\documentclass[runningheads]{llncs}
\usepackage[T1]{fontenc}
\usepackage{xcolor}

\usepackage[misc,geometry]{ifsym}
\usepackage{graphicx}
\usepackage{microtype}
\usepackage{amsmath}
\usepackage{algpseudocode}
\usepackage{algorithm2e}
\usepackage{comment}
\begin{document}
\title{STPA-driven Multilevel Runtime Monitoring \\
for In-time Hazard Detection} %
\titlerunning{STPA-driven Multilevel Runtime Monitoring}
\author{Smitha Gautham$^\text{\Letter}$\inst{1} %
\and
Georgios~Bakirtzis\inst{2}\and
Alexander Will\inst{1}\and
Athira~Varma~Jayakumar\inst{1}\and
Carl R. Elks\inst{1}}
\authorrunning{Gautham et al.}
\institute{Virginia Commonwealth University, Richmond, VA USA\\
\email{\{gauthamsm,willar,jayakumarar,crelks\}@vcu.edu} \and
The University of Texas at Austin, Austin, TX USA\\
\email{bakirtzis@utexas.edu}}
\maketitle              %
\begin{abstract}

Runtime verification or runtime monitoring equips safety-critical cyber-physical systems to augment design assurance measures and ensure operational safety and security. Cyber-physical systems have interaction failures, attack surfaces, and attack vectors resulting in unanticipated hazards and loss scenarios. These interaction failures pose challenges to runtime verification regarding monitoring specifications and monitoring placements for in-time detection of hazards. We develop a well-formed workflow model that connects system theoretic process analysis, commonly referred to as STPA, hazard causation information to lower-level runtime monitoring to detect hazards at the operational phase. Specifically, our model follows the DepDevOps paradigm to provide evidence and insights to runtime monitoring on what to monitor, where to monitor, and the monitoring context. We demonstrate and evaluate the value of multilevel monitors by injecting hazards on an autonomous emergency braking system model.

\keywords{dynamic safety management, cyber-physical systems, STPA, runtime verification, runtime monitors, hazard analysis.}
\end{abstract}

\section{Introduction}

Cyber-physical systems (CPS) are increasingly challenging to assess at design time with respect to system errors or hazards that could pose unacceptable safety risks during operation~\cite{redfield2017verification}. These challenges lead to the need for new methods allowing for a continuum between design time and runtime or operational assurance~\cite{fremont:2021}. Safety and security assurance at design level must be extendable to the runtime domain, creating a shared responsibility for reducing the risk during deployment. These emerging methods include dynamic safety management~\cite{trapp2018towards}, DepDevOps (dependable development operations continuum)~\cite{bruel_towards_2020}, systematic safety and security assessment processes such as STPA (system-theoretic process analysis) and STAMP (systems-theoretic accident model and processes), and MissionAware~\cite{bakirtzis2017mission}.

One emerging solution to help with the DepDevOps continuum is runtime monitoring or verification that observes system behavior and provides assurance of safety and security during the operational phase~\cite{bruel_towards_2020,leucker2009brief}. Runtime verification uses a monitor that observes the execution behavior of a target system. A monitor is concerned with detecting violations or satisfactions of properties (e.g., safety, security, functional, timeliness, to name a few) during the operation phase of a CPS. Execution trace information (i.e., states, function variables, decision predicates, etc.) is extracted directly from the CPS and forwarded to the monitor, where temporal logic expressions, called critical properties, are elaborated with this trace data for an on-the-fly verification of system behavior. %

To have effective runtime monitors, identifying critical properties to detect hazards (\emph{what} to monitor) and efficiently placing monitors where hazards may originate (\emph{where} to monitor) is crucial. However, most runtime monitoring frameworks for CPS emphasize \emph{how to monitor}~\cite{sanchez2019survey}. That is, runtime monitoring languages and tools primarily focus on (1) the expressiveness of the runtime verification language to capture complex properties, and (2) instrumenting a system to extract traces for monitoring, assuming the \emph{what to monitor} comes from some higher-level safety analysis process or methodology. Integrating system-level hazard analysis processes with runtime monitor design is essential for ``end-to-end'' functional safety assessment standards such as IEC-61508 and ISO 26262 that require traceable safety assurance evidence from requirements to design to implementation.

\vspace{.5em}
\noindent
\textbf{Contributions.} \quad
Our paper develops a well-formed workflow model which connects STPA hazard analysis information to lower level runtime monitoring used to detect hazards at the operational phase. Specifically, our model follows the DepDevOps paradigm to provide evidence and insights to runtime monitoring on: (1) what to monitor, (2) where to monitor, and (3) the context of the monitoring. Our work addresses the gap between safety analysis and runtime monitor formulation. 

In particular, we simulate hazard scenarios specified by STPA using model-based design and engineering (MBDE) tools, in our case MathWorks Simulink, to understand the boundary where a system can transition from a safe into an unsafe state. During hazard analysis, simulating hazard scenarios can reveal losses and their causal factors. We can thereby design well-informed context-aware runtime monitors to augment verification and validation (V\&V) performed at design time. 

\vspace{.5em}
\noindent
\textbf{Related Work.} \quad
STPA has been used extensively in avionics and automotive applications to study unsafe interactions among system components and how such interactions can result in unsafe control actions (UCAs) that may lead to system failures~\cite{leveson:2018}. STPA indicates that a UCA may result from multiple causal factors at different layers in a CPS. For efficient detection of these causal factors, we developed a multilevel runtime monitoring framework to support \emph{in-time anomaly detection}. In-time detection is the ability to detect hazard states before they lead to an accident and provide time for mitigation of the hazard. Multilevel monitoring was inspired by the fact that there is no single monitor type to solve in-time hazard detection problems of CPS. Instead, several types of monitors are usually needed to address this challenge~\cite{goodloe2010monitoring}.

STPA-driven runtime monitor design to ensure safety (and security) during the operational phase is an important and emerging research area. STPA is used to analyze unsafe system contexts in medical CPS to develop runtime safety monitors~\cite{ahmed_synthesis_nodate,zhou2021data}. In addition, work in the runtime monitoring domain emphasizes accuracy and integration over formal property development, whether by monitoring CPS~\cite{schwenger2020monitoring} or adding safety checking to a pre-existing system, such as monitoring distributed algorithms~\cite{liu2020assurance}. Properties for autonomous vehicle monitoring are derived from analyzing prior test results rather than being developed during the design process~\cite{zapridou2020runtime}. Our work, instead, intends to integrate runtime verification into CPS by creating properties through hazard analysis built into system design.

Service-oriented component fault trees are used for property derivation for runtime monitors with safety in mind~\cite{reich2020engineering}. Runtime monitors focus on the fault-tolerant qualities~\cite{haupt2019runtime} rather than emphasizing property generation, whereas property generation is our primary focus. Design-time safety measures that use STPA and model-based system engineering similar to our autonomous emergency braking (AEB) case study could incorporate our methods for runtime assurance~\cite{duan2022}. Attacks occur in hardware, communication, and processing levels within complex systems~\cite{cui_review_2019}, and using monitors at multiple system levels can increase causal factor awareness~\cite{daian_rv-ecu_2016,gautham_multilevel_2019,gautham_multilevel_2020}.

\section{STPA-driven Runtime Monitor Design}
\label{sec:frame}
\begin{figure}[!t]
\centering
\includegraphics[width=.9\textwidth]{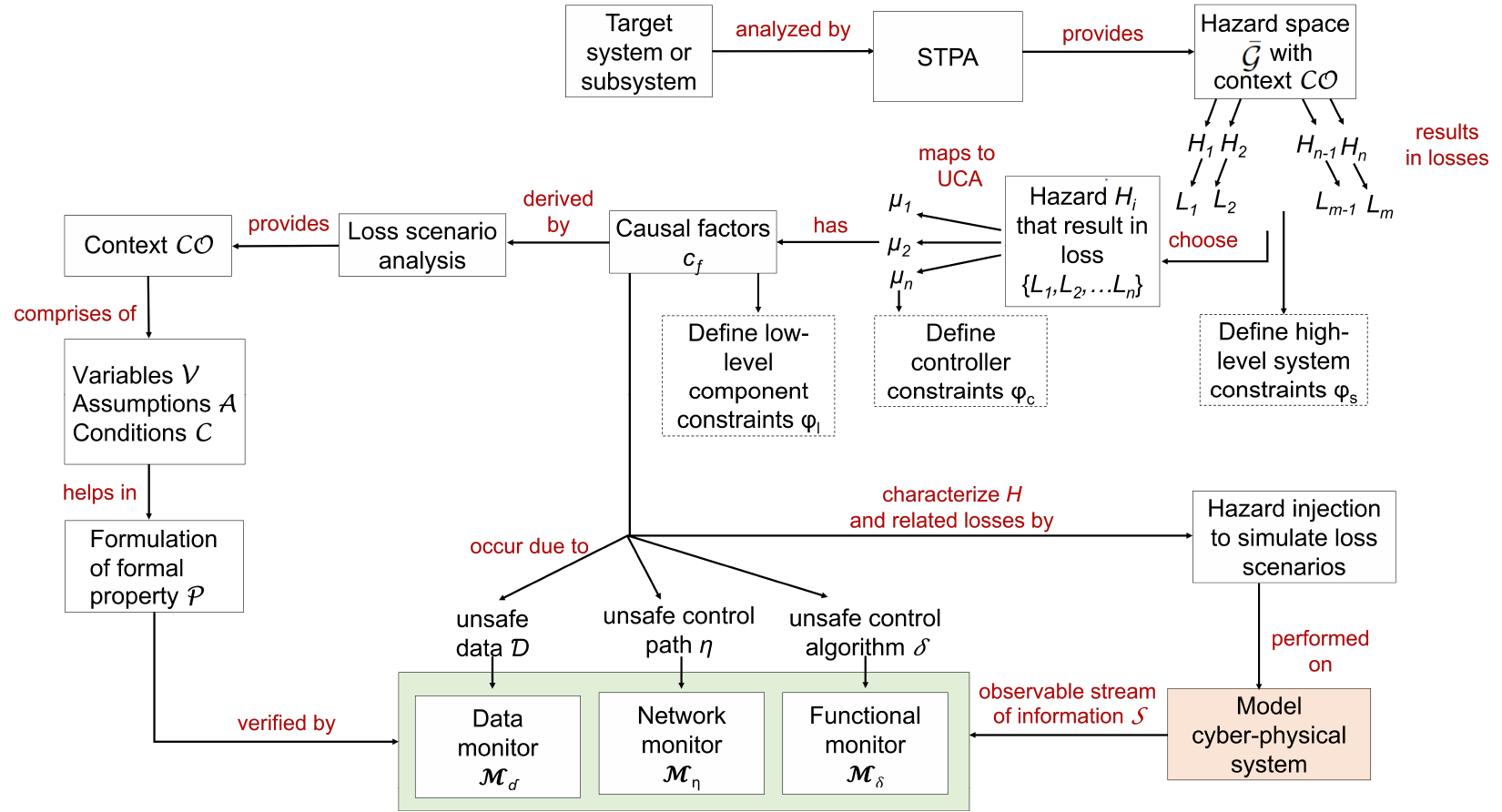}
\caption{STPA-driven runtime monitor generation.} 
\label{ff}
\end{figure}

An important motivation for this work is to explore an integrative approach to in-time hazard detection and informed risk that incorporates system level analysis into the design of monitoring architectures. 
Accordingly, we develop a STPA-driven model-based process for identifying and simulating hazard scenarios for designing multilevel runtime monitors (Fig.~\ref{ff}). 

\subsection{Losses, Hazards and Unsafe Control Actions}

A CPS consists of multiple coordinating components, continuously sensing and processing inputs from the physical domain and human users, and performing software-intensive tasks to produce time-critical outputs. This complex interaction among system components at specific system states increases the possibility of transitioning a system from a safe operating region $\mathcal{G}$ into an unsafe hazard space $\mathcal{\bar{G}}$. We denote all the identified hazards as $H =\{H_{1}, H_{2}, H_{3} \cdots H_{n}\}$. Such hazards can result in losses $L =\{L_{1}, L_{2}, L_{3} \cdots L_{m}\}$ that include loss of life, damage to property, to name a couple. Higher-level safety constraints (system constraints) $\varphi_s$ are derived from hazard analysis. These safety constraints result in safety requirements $R_{s}$ that inform the system development stage.

We denote the finite set of all possible control actions with $\Sigma = \mu \cup \alpha$ and denote the set of unsafe and $\alpha$ the set of safe control actions with $\mu$. UCAs $\mu$ can drive the system to a hazardous state $\mathcal{\bar{G}}$. Every specific hazard $H_{i}$ can be related to a finite subset of UCAs denoted by $u_{k}$, where $u_{k} \subseteq \mu$. The context, $\mathcal{CO}$, determines if an action is safe or unsafe. For example, a braking action at a given time $t$ may be safe to avoid a collision. Whereas, the same braking action may be unsafe on snowy road conditions as it may be unable to mitigate a collision due to delayed braking. An earlier braking action or a collection of actions may be needed for to be safe in a particular scenario.

Safety constraints $\varphi_c$ (sometimes called controller constraints) and safety requirements $R_{c}$ are defined at the controller level for all $\mu_i \in \mu$. Although $\varphi_c$ are typically incorporated into a design to prevent a hazard, there can be faults/attacks during operation that can violate the safety requirements $R_{c}$ imposed by the designer. Furthermore, in some scenarios, $\varphi_c$ cannot be enforced in a system. Runtime safety assurance via monitors is important for promptly detecting safety constraints and requirements violations to prevent a hazard.

\subsection{Causal Factors and Relation to Multilevel Monitoring }

Finding the possible causes for a specific UCA $\mu_i \in  u_i$ is an essential step in preventing a hazard $H_i$. When a violation is detected, providing a timely safe control action $ \alpha_i$ can prevent a system from transitioning into the unsafe operating region  $\mathcal{\bar{G}}$,  consequently avoiding a hazard.  

We denote the causes for a UCA $\mu_i$ as a causal factor $c_{f}$. Causal factors $c_{f}$ are directly related to a given UCA $\mu_i \in u_i$ (Fig.~\ref{ff}), where in a given context $\mathcal{CO}$ a causal factor $c_{f}$ causes the UCA $\mu_i$ and may lead to the associated hazard $H_{i}$. To determine causal factors $c_{f}$ for each UCA $\mu_i$ we define loss scenarios, which reveal the context $\mathcal{CO}$ in which hazard $H_{i}$ may occur. The context has a set of variables $\mathcal{V}$ which can take multiple values depending on the system state or environment or vehicle conditions, a set of assumptions $\mathcal{A}$ made on certain variable values, and a set of system conditions $\mathcal{C}$ of a system based on the variables and assumptions~\cite{thomas:2013}. A unique combination of deviation in values for the variables $\mathcal{V}$ with a violation of assumptions $\mathcal{A}$ related to a condition $\mathcal{C}$ forms the basis for a causal factor $c_{f}$ for a hazardous control action $\mu_i$. Thus, the context can be expressed as a mapping $\mathcal{CO}\colon\mathcal{V} \times \mathcal{A} \times \mathcal{C} \to c_{f}$. Once causal factor analysis is complete, low-level component constraints  $\varphi_{l}$ are generated to define the boundary for safe operation at the component level. Components can be both hardware and software, i.e. functional modules such as controllers and other subsystems such as communication buses, sensors etc. in a CPS. Fault/hazard injection approaches are used to strategically inject faults to simulate the deviation in $\mathcal{V}$, $\mathcal{A}$, and $\mathcal{C}$ to create loss scenarios and test the boundaries of these constraints. 

Further, the causal factors can specifically be related to one of the levels or layers in a multilevel view of the system. STPA provides suggestions for classification of causal factors for hazards that can occur at multiple levels, including controller-based (inadequate control algorithm, flawed control algorithm), input-based (unsafe data from other controllers, failure of sensor inputs), and control path-based (network delays, flaws in data process algorithm in a controller)~\cite{leveson:2018}. For our multilevel monitoring structure, we define the following levels: unsafe data $\mathcal{D}$, unsafe processing $\delta$, and unsafe behavior in the communication path $\eta$. The causal factors related to unsafe inputs to a controller from sensors, user inputs, or input from another controller as well as unexpected/incorrect data patterns are $\mathcal{D} = \{d_{1}, d_{2}, \cdots, d_{n}\}$, where $\mathcal{D} \in \mathcal{V}\times\mathcal{A} $. The causal factors related to flaws in the control algorithm and incorrect functional behavior in the controller are $\delta =\{\delta_{1}, \delta_{2},\cdots,\delta_{n}\}$, where $\delta \in \mathcal{V}\times\mathcal{A}\times\mathcal{C}$. The causal factors related to flaws in the control path through which inputs/outputs are communicated between the subsystems are $\eta=\{\eta_{1}, \eta_{2}, \cdots,\eta_{n}\}$, where $\eta \in \mathcal{V}\times\mathcal{A}\times\mathcal{C}$. For timely detection of such causal factors before they result in a UCA $\mu_i$, we believe that a viable approach is to employ monitors at these various levels of processing and integration where the vulnerabilities originate.

\subsection{Multilevel Runtime Monitoring Framework}

Multi-level monitoring extends traditional runtime verification or monitoring by providing a monitor classification or organization schema that maps monitor types to various functions or components in distributed real-time architectures~\cite{goodloe2010monitoring}. In this work, we augment a multi-level monitoring framework~\cite{gautham_multilevel_2020} as it directly addresses monitoring CPS from multiple layer perspectives.
A monitor $\mathcal{M}_{a}$ observes streams of time stamped information from a target CPS. A \emph{stream}, denoted as $\mathcal{S}_{a} = {\mathcal{S}_{a}(t - m), \cdots, \mathcal{S}_{a}(t - 2), \mathcal{S}_{a}(t - 1),  \mathcal{S}_{a}(t)}$, where $\mathcal{S}_{a}$, is a sequence of time-stamped information, from the past $m$ instances starting with $\mathcal{S}_{a}(t - m)$ and ending at the current instance $\mathcal{S}_{a}(t)$. The $a$ subscript denotes a stream associated with a specific part of the system. We denote the set of all streams from different parts of a CPS as $\mathcal{S}$, in particular, $\mathcal{S}_{a} \in \mathcal{S}$ for all streams $\mathcal{S}_{a}$.

The streams of information from the CPS that we want to verify as being compliant to safe operation requirements can be represented as a monitorable property $\mathcal{P}$ derived from %
component constraints $\varphi_{l}$ after STPA analysis. The property $\mathcal{P}$, also referred to as a monitor specification, is a checking condition that represents the conditions given by a context $\mathcal{CO}\subseteq\mathcal{V} \times \mathcal{A} \times \mathcal{C}$ (Fig.~\ref{ff}), and is most often expressed in temporal logic. Thus, in multi-level monitoring, a monitor of a specific type placed at a specified level detects unsafe or hazardous conditions for the stream it is observing. We classify monitors (and their associated properties) as data, network, or functional monitor types depending on the causal factors $c_f$ and the possible location of emerging hazard states given by STPA. We consider the following three types of monitoring for CPS: input-output (I/O) data-oriented monitors of type $M_d$, network-oriented monitors of type $M_\eta$, functional monitors of type $M_\delta$.

\begin{itemize}
\item \textbf{Data Monitor $\mathcal{M}_{d}$} observes streams of data from sensors and actuators that provide an interface to the physical environment, signals behavior of a controller and verifies the data integrity. The causal factors related to $\mathcal{D}$ , i.e. unsafe input from sensors or from other controllers are verified by $\mathcal{M}_{d}$. 
\item \textbf{Network Monitor $\mathcal{M}_{\eta}$} verifies the integrity of the data received by the communication layer by observing streams of information from the network layer. They check for signal faults, incorrect signaling protocol, timing delays etc. They observe causal factors related to $\eta$, i.e. unsafe control path. 
\item \textbf{Functional Monitor $\mathcal{M}_{\delta}$} verifies properties for the system's functional behavior. For example, the relation between input and output of a controller is verified by a functional correctness property.  In particular, $\mathcal{M}_{\delta}$ observe causal factors related to $\delta$, i.e., an unsafe control algorithm, by observing streams of information consisting of system states, internal variables, memory read/writes, and event counts.
\end{itemize}

\section{Monitoring an AEB Controller}

\begin{figure}[!t]
\centering
\includegraphics[width=.75\textwidth]{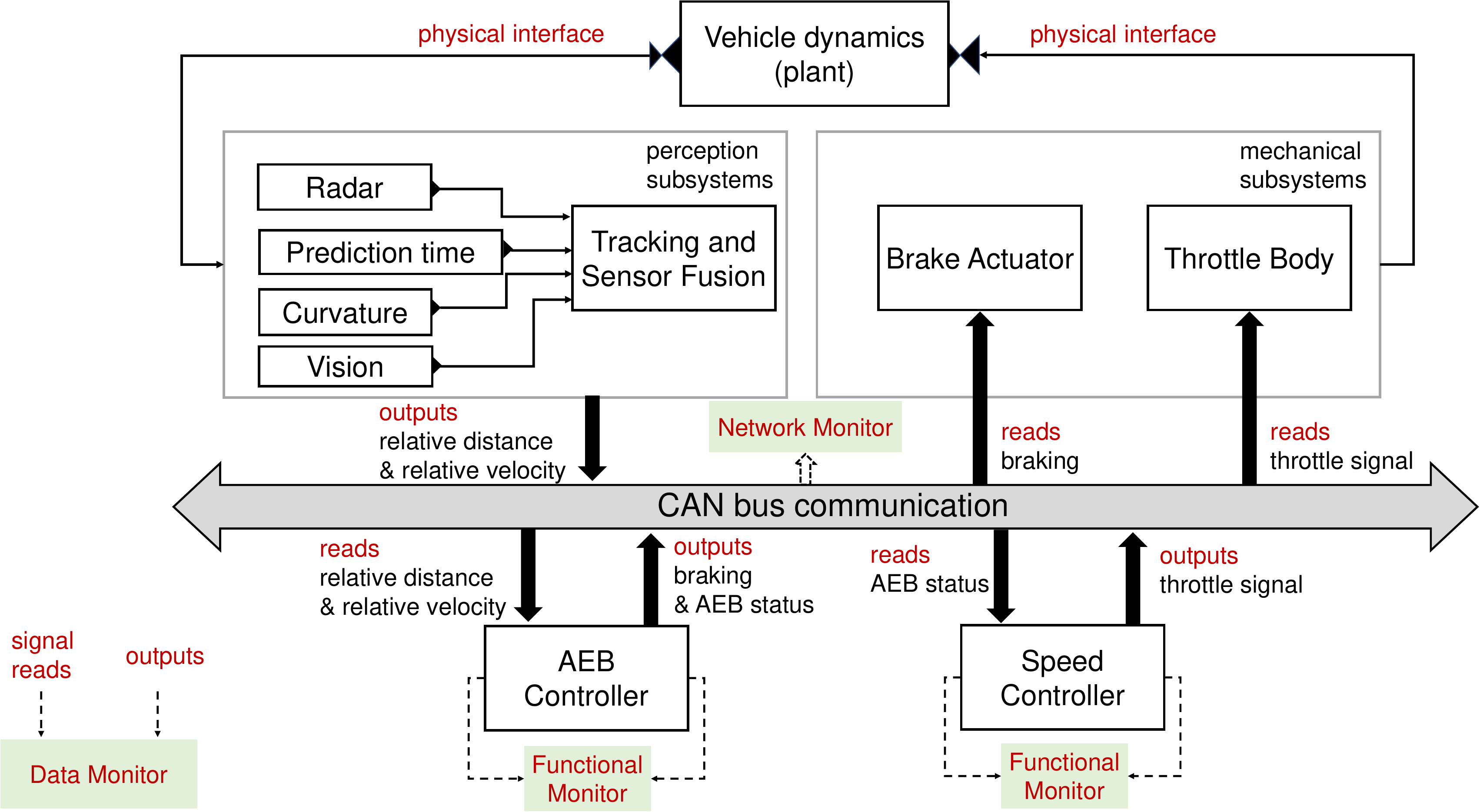}
\caption{Schematic of an autonomous emergency braking (AEB) system.} 
\label{aeb}
\end{figure}

A simplified AEB system model~\cite{mathworks:2021} is a representative system for studying the methodology for STPA-driven runtime monitor design (Fig.~\ref{aeb}). The output of the AEB controller determines the braking state that decelerates the ego car, which is a car with autonomous features. 

A model of the vehicle dynamics module was considered whose output---together with the scenario under consideration---determines the inputs to the radar and vision sensors. The outputs of these sensors are fused to estimate the relative distance and relative velocity between the ego car and the ``most important object'' (MIO). The MIO is not always the lead car. For example, if a pedestrian comes in front of the ego car, this would be the MIO. Based on these inputs (distance and velocity relative to the MIO), the AEB controller estimates the braking state (Fig.~\ref{aeb}). When the ego car is at a safe distance but gets closer required for safe operation, an alert, forward collision warning, is issued. If the driver does not brake or the braking is insufficient, then the AEB engages the ``stage I'' partial braking (PB1) at a certain critical relative distance. If this does not suffice, ``stage II'' partial braking (PB2) is applied at a closer relative distance, and then full braking (FB) is engaged. This action decelerates the car to avoid a collision characterized by a minimum headway distance when the velocity of the ego car reaches zero. Runtime monitors of data, network and functional types are placed at different levels in a CPS (Fig.~\ref{aeb}).

\subsection{STPA for AEB}

\textbf{Losses and Hazards.}\quad From our analysis, we consider the losses $L$ that must not occur, and the hazards $H$ related to the losses $L$ are described below. These form the foundation for producing UCAs (Fig.~\ref{De}). For some of the hazards, we mention sub-hazards to cover different cases. Some illustrative subsets of losses and hazards:

\vspace{.5em}\noindent
\begin{tabular} {p{1cm}l}
L-1	& Loss of life or injury due to collision \\
L-2	& Loss via damage to the vehicle or property (repair, fines etc.)\\
L-3 &	Loss of reputation
\end{tabular}

\vspace{.5em}\noindent
\begin{tabular} {p{1cm}l}
H-1	& Unsafe headway distance to the MIO [L-1, L-2, L-3]\\
H-1.1 & Unsafe headway distance to vehicles, pedestrians [L-1, L-2, L-3]\\
H-1.2 & Unsafe headway distance to sidewalks, curb etc. [L-2, L-3]\\
H-2 & Vehicle is traveling at an inappropriate speed. [L-1, L-2, L-3]
\end{tabular}
\vspace{.5em}

Higher-level system constraints $\varphi_{s}$ are derived from rephrasing of the hazard statements as a binding mandatory requirement.  For example, the system constraint for the hazard H-1 is: ``$SC^1_\mathrm{system}$ ($\varphi_{s}$) The Ego car must always maintain a safe distance to the MIO.''

\begin{figure}[!t]
\centering
\includegraphics[width=.7\textwidth]{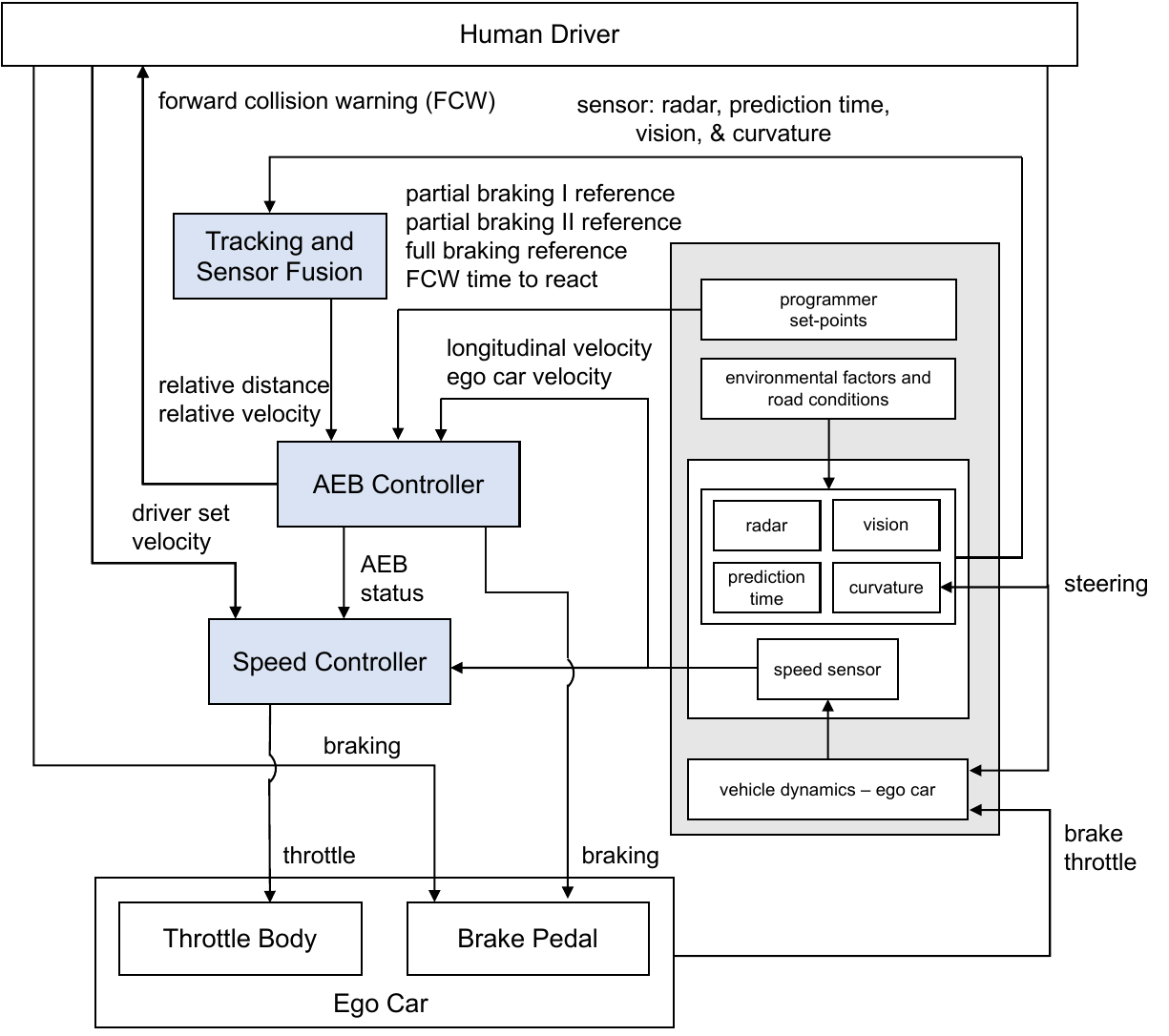}
\caption{STPA control structure diagram for AEB system.} 
\label{cs}
\end{figure}
\vspace{.5em}\noindent\textbf{Control Structure Diagram.}\quad The STPA control structure diagram shows all the components in the AEB system along with vehicle dynamics and environmental factors. It is a hierarchical control structure with a human driver at the top, brake and throttle controllers in the middle, and mechanical components such as the throttle body and brake pedal at the bottom of the diagram.

Next, we identify UCAs $\mu$ that can occur in the AEB system. This step occurs after loss and hazard determination because UCAs $\mu$ directly cause hazards (Fig.~\ref{De}). The AEB controller provides a \emph{braking} signal to the brake pedal, an \texttt{AEBstatus} signal to the speed controller, and forward collision warning (FCW) to the driver (Fig.~\ref{cs}). \emph{Braking} is a deceleration signal with different braking levels PB1, PB2, and FB (Section 3). Whenever the AEB controller activates the brakes, the \texttt{AEBstatus} signal indicates to the speed controller the braking level applied. \texttt{AEBstatus} 1 indicates ``Partial Braking I'', \texttt{AEBstatus} 2 indicates ``Partial Braking II'', and \texttt{AEBstatus} 3 indicates ``Full Braking'', all as applied by the AEB. Based on the \texttt{AEBstatus}, the speed controller provides or ceases to provide an acceleration signal to the throttle.

\vspace{.5em}\noindent\textbf{Identifying Unsafe Control Actions (UCAs).}\quad 
Based on the control structure diagram analysis, we illustrate a subset of UCAs (Table~\ref{uca}). As an example, we state the controller constraint $\varphi_{c}$ for UCA 1:
``$SC_\mathrm{controller}^1$ ($\varphi_{c}$) AEB must provide a \emph{braking} signal when MIO is approaching the Ego car and AEB detects an imminent collision [UCA 1].'' Here, \emph{braking} and \emph{detection of imminent collision} are AEB's control actions. Finding incorrect or untimely control actions can guide designers towards finding comprehensive loss scenarios and low-level safety requirements for \emph{braking} and correct \emph{detection of imminent collision}.

\begin{table}[!t]
\caption{Partial List of Unsafe Control Actions in the AEB system.} 
\includegraphics[width=\textwidth]{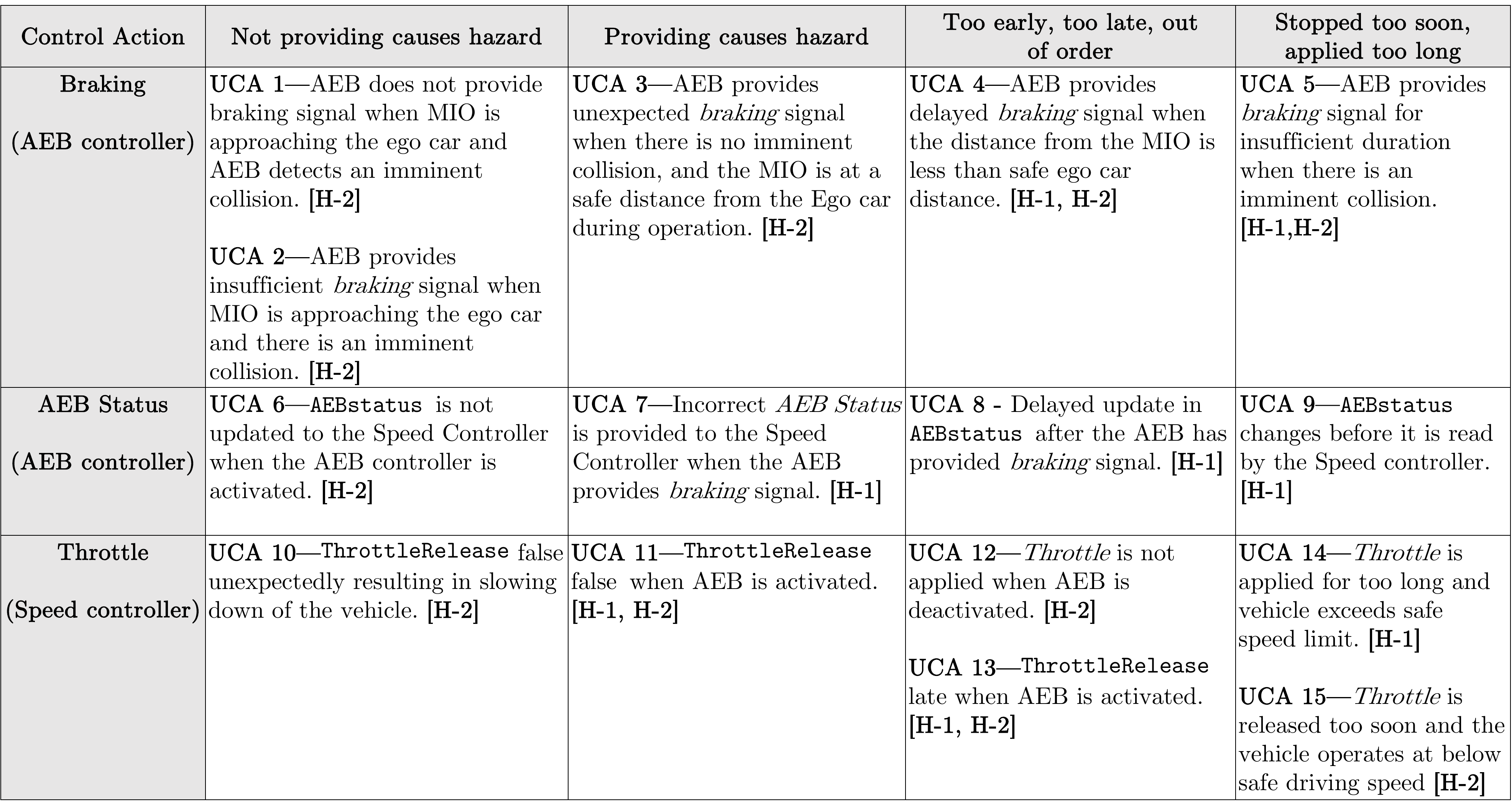}
\vspace{-2em}
\label{uca}
\end{table}

\subsection{Loss Scenarios and Causal Factors as Design Guides\\ for Multilevel Runtime Monitoring}

\begin{figure}[t!]
 \centering
 \includegraphics[width=\textwidth]{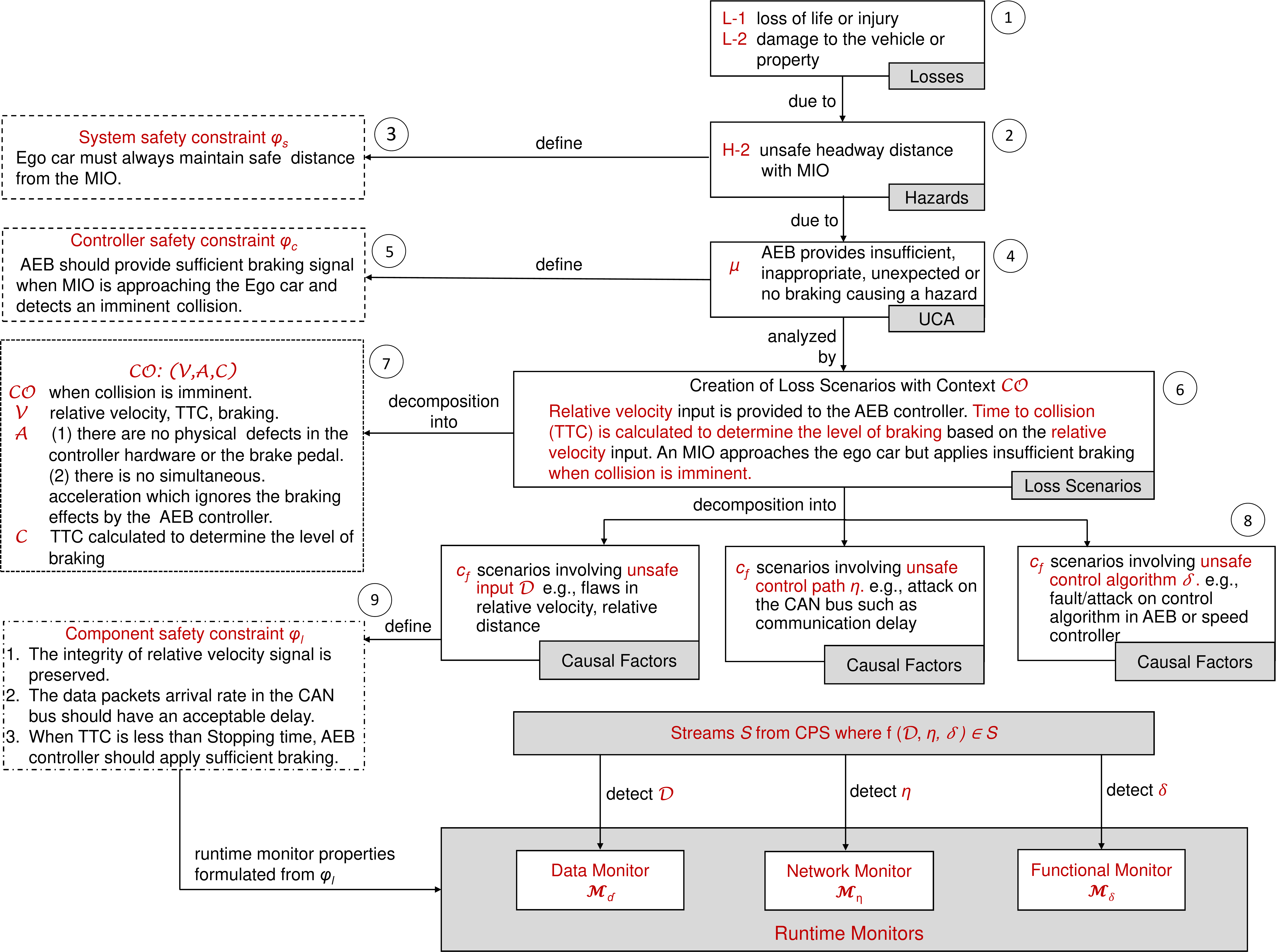}
\caption{Deriving multilevel runtime monitor properties from STPA for AEB system (numbers denote the order of the workflow). } 
\label{De}
\vspace{-2em}
\end{figure}

Causal factors for a UCA $\mu_i$ provide insights on complex subsystem interactions and failure patterns that are critical in developing component-level safety constraints $\varphi_{l}$. These low-level component constraints $\varphi_{l}$ are vital for detection of fault/attack and possible isolation of the causal factors. As a case study, we identify the potential causal factors that result in unsafe \emph{braking} by the AEB controller and unsafe throttle action by the speed controller. Causal factors for UCAs could be due to a) failures related to the controller, b) inadequate control algorithm, c) unsafe control inputs, and d) inadequate process model as described in the STPA handbook~\cite{leveson:2018}. To determine the causal factors, Fig.~\ref{De} explains that we must describe loss scenarios based on each UCA in $\mu$ as to both realize the context $\mathcal{CO}$ and formulate each of the different causal factors $c_f$. For the AEB system, we describe two scenarios which describe the context $\mathcal{CO}$ for unsafe \emph{braking} UCA. In each scenario we identify the component level safety constraints $\varphi_{l}$ based on the illustrated causal factor $c_f$ along with the runtime verification properties used to detect the causal factor $c_f$. We express the monitor properties using event calculus temporal formal language~\cite{shanahan:2009}. 

\vspace{.5em}
\noindent
\textbf{Scenario 1: Safe Braking Distance.} \quad
The vehicle is operating and begins approaching an MIO. The AEB applies \emph{braking} in accordance with its control algorithm and updates the \texttt{AEBstatus} to a non-zero number corresponding to the level of \emph{braking} applied. The speed controller applies throttle and ignores the change in \texttt{AEBstatus}. If the speed controller continues acceleration while \emph{braking} is occurring, the braking components experience undue strain and may fail, leading to potential unsafe headway [H-2] and collision [L-1]. The rationale for simultaneous \emph{braking} and acceleration from the speed controller’s perspective varies depending on the context $\mathcal{CO}$ in the scenarios. Potential causes include:

\begin{figure}[t!]
\centering
\includegraphics[width=.85\textwidth]{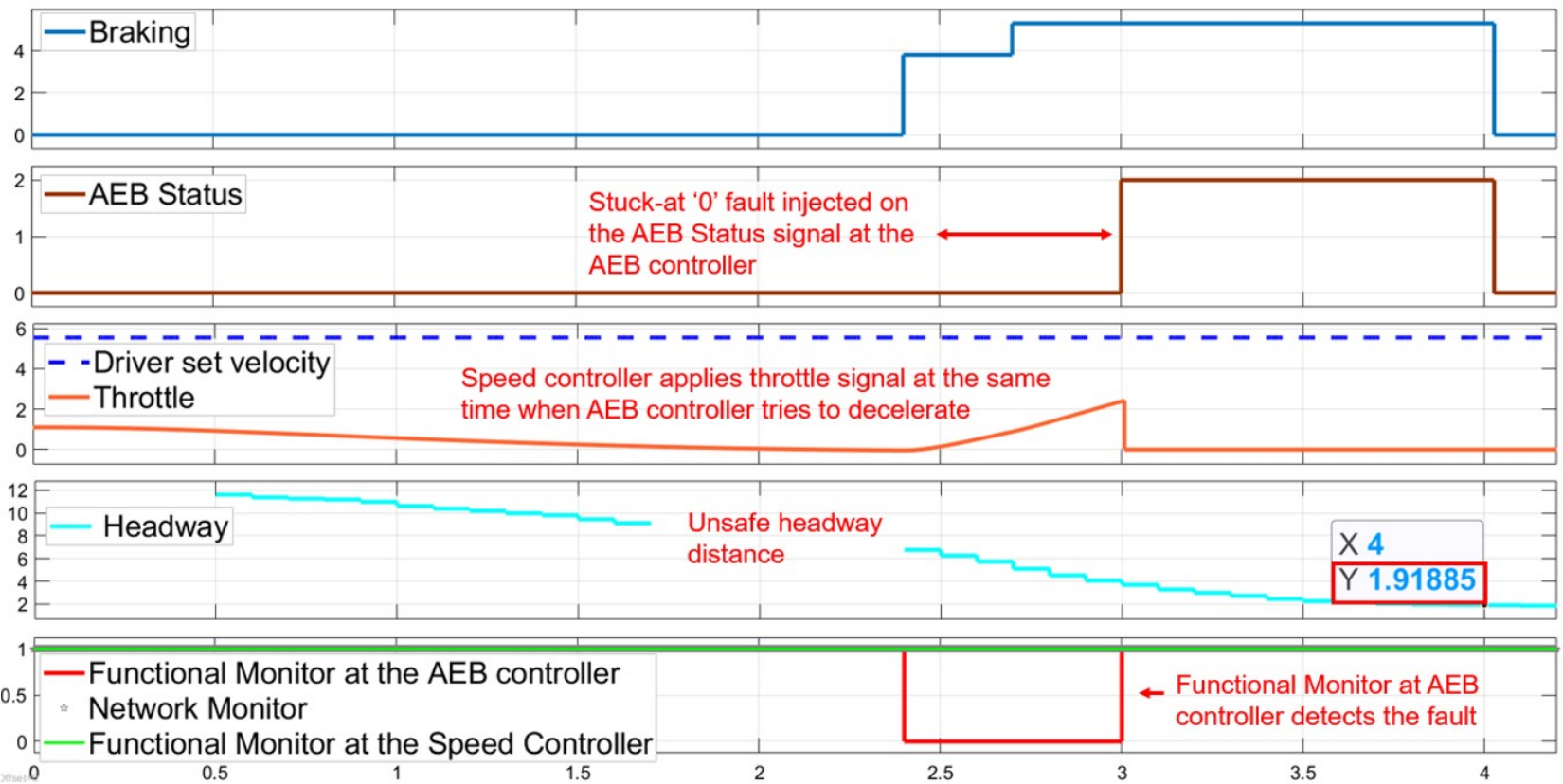}
\caption{Localized monitors at each level are beneficial (Scenario 1).} 
\label{sc1}
\end{figure}

\begin{description}
\item [Scenario 1a.] The speed controller has an inadequate control algorithm and does not release the throttle when the \texttt{AEBstatus} is non-zero.
\item [Scenario 1b.] The communication between the AEB and the speed controller is delayed. Thus, the speed controller is not aware of the change in \texttt{AEBstatus}, and it keeps the throttle on when the \texttt{AEBstatus} is non-zero.
\item [Scenario 1c.] The AEB does not properly update \texttt{AEBstatus} signal, even after beginning braking [UCA-6,7]. Thus, the speed controller believes the AEB is not \emph{braking} and keeps the throttle on when the \texttt{AEBstatus} is non-zero.
\end{description}
 The context in the scenario can be expressed as a function of $\mathcal{V}$, $\mathcal{A}$ and $\mathcal{C}$~\cite{thomas:2013}. For example in scenario 1, throttle and \texttt{AEBstatus} are the variables $\mathcal{V}$, the assumption $\mathcal{A}$ is that the input \texttt{AEBstatus} is accurate and the vehicle is in motion. ``Release throttle when \texttt{AEBstatus} is non-zero'' is the condition $\mathcal{C}$. 
The component-level safety constraints and their corresponding properties are listed below:

\vspace{.5em}\noindent
$SC_{\mathrm{component}}^1$ ($\varphi_{l}$)	The Speed Controller should release the throttle when the \texttt{AEBstatus} is non-zero.
\begin{property}[Detects inadequate control algorithm in speed controller (Scenario~1a)] ``If \texttt{AEBstatus} is equal to 1, 2, or 3, the throttle should be released.'' This ensures that the car throttle is not engaged when the brake is engaged by the AEB,
\[\textsf{Happens}(\texttt{AEBstatus} = 1 \vee 2 \vee 3, T)
\Rightarrow \textsf{HoldsAt}(\texttt{ThrottleRelease},T).\]
\end{property}

\noindent
$SC_{\mathrm{component}}^2$ ($\varphi_{l}$)	The data packet's arrival rate via the CAN bus should have an acceptable delay.
\begin{property}[Detects flaw in control path to speed controller (Scenario 1b)]
``The time interval between two successive packet arrival via the CAN bus should be less than $T_\mathrm{safe}$.'' This condition ensures that the consecutive packets $\mathrm{Packet}_{A}$ and $\mathrm{Packet}_{B}$ should arrive at time $T_{a}$ and $T_{b}$ respectively, where $T_{d}=T_{b}-T_{a}$ should satisfy the condition $T_{d}<T_\mathrm{safe}$, 
\(\textsf{Happens}(\mathrm{Packet}_A, T_a) \wedge \textsf{Happens}(\mathrm{Packet}_B,T_b).
\)
\end{property}

\noindent
$SC_{\mathrm{component}}^3$  ($\varphi_{l}$) When the AEB controller begins the \emph{braking} action, the \texttt{AEBstatus} should be updated accordingly.
\begin{property}[Detects inadequate control algorithm in AEB (Scenario 1c)]
``If deceleration is greater than $D_\mathrm{safe}$ $\mathrm{m}/\mathrm{s}^{2}$, the \texttt{AEBstatus} should be non-zero.'' This property ensures that the \texttt{AEBstatus} corresponds to the AEB controller's current braking signal,
\(\textsf{Happens}(\mathrm{Deceleration} > D_\mathrm{safe} , T) 
\Rightarrow \textsf{HoldsAt} (\texttt{AEBstatus} \neq 0,T). \)
\end{property}

  \begin{figure}[!t]
  \centering
\includegraphics[width=.85\textwidth]
{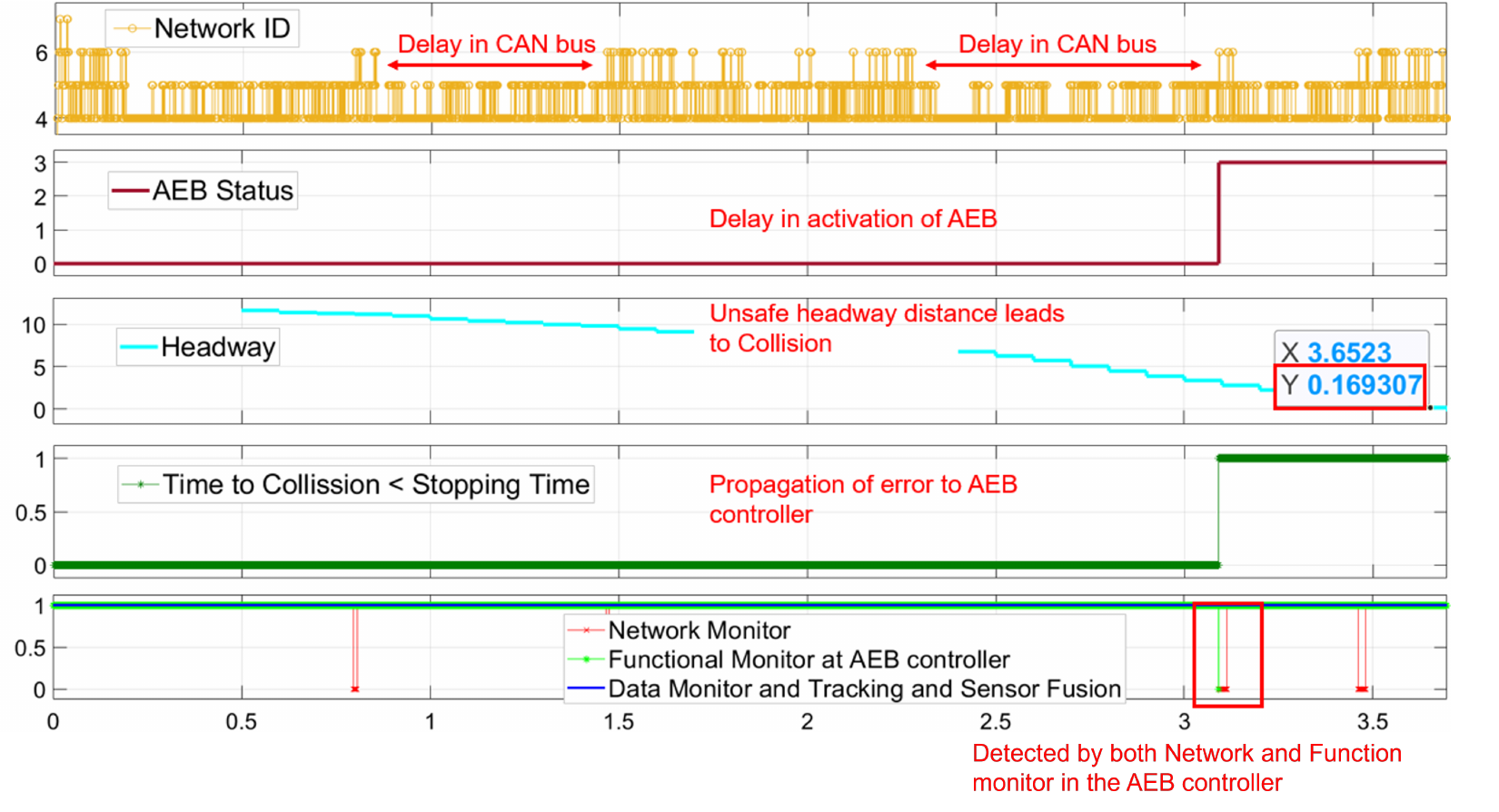}
\caption{Multilevel monitoring for in-time detection (Scenario 2).} 
\label{sc2}
\vspace{-1.5em}
\end{figure}

\noindent
\textbf{Scenario 2: Communication Delay.} \quad
The vehicle is operating and begins to approach an MIO. There is a communication delay in sending the relative distance and relative velocity signals from the Tracking and Sensor Fusion module to the AEB controller resulting in delayed calculation of Time To Collision (TTC). Because of this, sufficient and timely braking is not applied. The component level safety constraints and runtime property based on this scenario were formulated similar to Scenario 1b to detect delay in communication in the CAN bus.

\vspace{.5em}\noindent
\textbf{Hazard Injection and Monitor Detection.} \quad
Using a model-based fault injection toolbox~\cite{jayakumar2020property}, faults and attacks were injected strategically to simulate the loss scenarios 1 and 2. STPA provides a systematic method to analyze the system and identify loss scenarios. After identifying loss scenarios, we explore adequacy of the causal factor analysis by property-based hazard injection~\cite{jayakumar2020property}.

For the first scenario, faults were injected on the \texttt{AEBstatus} signal in the AEB controller (Scenario 1c). Although the AEB controller provides braking action, the speed controller is unaware of the braking and continues to apply throttle due to the fault. This results in simultaneous braking and acceleration of the vehicle, thus causing the unsafe headway distance hazard [H-1]. The headway reduces to 1.9 meters at 4 seconds (Fig.~\ref{sc1}) (safe headway distance should be at least 2.4 meters). The functional monitor at the AEB controller detects the fault much earlier than the occurrence of the hazard. This error is detectable only by having a localized monitor at the AEB controller. The functional monitor at the speed controller and the network monitor do not detect such a fault, as a fault on \texttt{AEBstatus} neither changes the functionality of the speed controller nor the network behavior. Thus, having local monitors at each level is beneficial in \emph{early detection} and \emph{isolation of faults/attacks}. 

For the second scenario, we emulate a malicious node attack where sporadic messages on the CAN bus causes delay in the communication of packets between the tracking and sensor fusion module and the AEB controller. The network-level monitor detects this scenario (Fig.~\ref{sc2}). A message injection attack at the network layer also results in violation of a functional property in the AEB controller  which verifies the control algorithm at the AEB controller level. The AEB controller decides on the level of braking based on ``time to collision'' and ``stopping time'' (time from starting braking to coming to a complete stop). An attack on the network layer results in violation of the functional property \emph{when $\mathrm{TTC} < \mathrm{stopping time}$, ego car velocity should be decreasing}, thus demonstrating error propagation from one layer to another. While our simulation example confirms that the error is caught both by the network monitor and the functional monitor, the simulation shows that the network monitor detects the error much earlier than the functional monitor at the AEB controller. This use case scenario validates the in-time early detection of the emerging hazard before error propagation reaches the system's output boundaries. In fact, when a property violation goes unnoticed at one level, they are often detected by another monitor in the hierarchy as effects propagate, thus \emph{improving hazard detection coverage}.

The workflow integrates requirement elicitation through STPA into the direct creation of runtime monitors by decomposing the causal factors at different system levels on the basis of component safety constraints (Fig.~\ref{De}).  There is an iterative feedback for refinement of safety constraints after hazard injection.

\section{Conclusion}
\label{sec:conc}

We developed an integrative approach to in-time hazard detection that incorporates system-level analysis into the design of runtime monitoring architectures. Integrative approaches to runtime monitoring for hazard detection in CPS are needed to augment the technical basis for DepDevOps style methods. %
We demonstrated that the systematic nature of STPA hazard analysis is beneficial in deriving and refining multilevel monitoring properties related to causal factors. By developing monitors across multiple system levels, we can accurately detect the origin of a hazard even when it propagates errors across different CPS levels.

In other words, when faults go undetected at their original location, monitors at other system levels can detect propagated errors, thus increasing hazard detection coverage. We also found that MBDE methods and tools significantly improve the productivity of STPA and assist in evaluating runtime monitoring schemes for hazard coverage and refinement.

\bibliographystyle{splncs04}
\bibliography{manuscript}
\end{document}